\documentclass{emulateapj}
\usepackage{apjfonts}
\usepackage{times}
\usepackage{epsfig}
\def\apj{ ApJ}

\def\mnras{MNRAS}

\def\apjs{ApJ Supp}

\shorttitle{CE and enthalpy}
\shortauthors{Ivanova \& Chaichenets}

\begin{document}

\title{Common envelope: enthalpy consideration}
\author{N.\ Ivanova$^1$, S.\ Chaichenets$^2$}
\altaffiltext{1}{University of Alberta, Dept. of Physics, 11322-89 Ave, Edmonton, AB, T6G 2E7, Canada}
\altaffiltext{2}{University of Alberta, Dept. of Mathematical and Statistical Sciences, CAB, Edmonton, AB, T6G 2G1, Canada}

\begin{abstract}{
In this {\it Letter} we discuss a modification to the criterion for the common envelope (CE) 
event to result in envelope dispersion.
We emphasize that the current energy criterion for the CE phase is not sufficient for an instability of the CE,
nor for an ejection.
However, in some cases, stellar envelopes undergo {\it stationary mass outflows},
which are likely to occur during the slow spiral-in stage of the CE event.
We propose the condition for such outflows, in a manner similar to the currently standard 
$\alpha_{\rm CE}\lambda$-prescription but 
with an addition of $P/\rho$ term in the energy balance equation, accounting therefore 
for the {\it enthalpy} of the envelope rather than merely the gas 
internal energy. This produces a significant correction,
which might help to dispense with an unphysically high value of
energy efficiency parameter during CE phase, currently required in the 
binary population synthesis studies to make the production 
of low-mass X-ray binaries (LMXBs)  with a black hole companion to  match the observations.
}
\end{abstract}


\section{Introduction}

The most important event in the formation and evolution of most close interacting binaries is the 
so-called common envelope (CE) event, during which the components of a binary system 
are engulfed by a common gaseous envelope, and the resulting interaction dramatically
shrinks their orbit \citep{BKS,Os76,Pa76}. 
Depending on the envelope structure and companion masses, the envelope is ejected to leave behind a close binary,
or the two stars merge. 
Dividing the parameter space into binaries that survive CE and those that do not, 
and determining the final separation for the former, is critical for calculating formation rates of low mass X-ray binaries (LMXBs), 
$\gamma$-ray bursts, as well as for LISA and LIGO events \citep{Bel08}.

In the standard treatment of CE outcomes,
the final separation of the binary is determined via ``energy formalism'' \citep{Webbink84,Livio88}, in which the binding energy of 
the (shunned) envelope is equated to the decrease in the orbital energy $E_{\rm orb}$:

\begin{equation}
E_{\rm bind} = E_{\rm orb,i} - E_{\rm orb,f} = -\frac{ G m_1 m_2} {2 a_{\rm i}} + \frac{ G m_{1\rm,c} m_2} {2 a_{\rm f}} 
\end{equation}
Here  $a_{\rm i}$ and $ a_{\rm f}$ are the initial and final binary separations, $m_1$ and $m_2$ are the initial star masses
and $ m_{1\rm,c}$ is the final mass of the star that lost its envelope.

$E_{\rm bind}$ is assumed to be the energy expense needed to remove the envelope to infinity
and is commonly adopted to be the sum of the potential energy of the envelope and its internal energy.
To characterise the donor envelope central concentration and simplify calculations, specifically for population synthesis,
a parameter $\lambda$ was introduced:

\begin{equation}
E_{\lambda, \rm bind} = - \int_{\rm core}^{\rm surface} \left ( \Psi(m) + \epsilon (m) \right) dm  = \frac {G m_1 m_{1,\rm e}} {\lambda R_1} 
\label{lambda}
\end{equation}
Here $m_{1,\rm e}$ is the mass of the removed giant envelope and $R_1$ is the radius of the giant star at the onset of CE,
$\epsilon$ is the specific internal energy and $\Psi(m)=-Gm/r$ is the  (gravitational) force potential (or specific potential energy). 
$E_{\lambda, \rm bind}$ can be found directly from stellar structure for any accepted core mass.

Another parameter, $\alpha_{\rm CE}$, is introduced as a measure of the energy transfer efficiency from the orbital energy 
into envelope expansion, and the balance of energy is written as

\begin{equation}
\alpha_{\rm CE} {\lambda} \left ( \frac{ G m_{1\rm,c} m_2} {2 a_{\rm f}} -\frac{ G m_1 m_2} {2 a_{\rm i}} \right ) =
\frac {G m_1 m_{1,\rm e}} {R_1}
\label{allam}
\end{equation}
Many authors choose to accept $\alpha_{\rm CE} \lambda \approx 1$, since for many (at least low-mass) stars $\lambda=1$, 
as can be found from detailed stellar structure, and $\alpha_{\rm CE}$ is bounded above by 1. 

The simplicity of the standard prescription resulted in its popular use 
in the binary population synthesis  calculations. However, once
accurate values of $\lambda$ from stellar structure calculations were determined,
this approach has shown inconsistencies with the observations, especially large for the formation of 
black hole LMXBs: in massive giants $\lambda \ll 0.1$  \citep{podsi03}, and it has been shown that   
with  $\alpha_{\rm CE}\le 1$ only an intermediate-mass companion could  avoid 
a merger; this challenges the formation of a low-mass X-ray binary with a low-mass companion in general \citep{Justham06}.

Here, we revise the energy requirements necessary to force a common envelope to disperse  
and discuss its possible application for LMXBs formation.


\section{Enthalpy considerations}\label{enthalpy}

\subsection{Total energy and instability}

While the definition of the gravitational binding energy is unequivocal, \citep[e.g.,][]{Chandra39}, 
there exists no authoritative source defining what expression should be used as the binding energy 
of a gaseous sphere in the sense of \S~1. To clarify, in treating the common envelope problem we are interested 
in the additional energy to be deposited in the envelope in order to disperse material to infinity. 
The `industry standard' at the moment is to use the total of the gravitational $U$ and the gas thermal $E$ 
energies, which, as we shall argue in this section, is not correct. 

As an illustrative limiting case, consider a star with the positive total energy\footnote{
	with zero of energy defined for all material evacuated to infinity, and pressure set to zero
} 
$W_0|_{t=0}>0$ to begin with, which is {\it ``kinetically stable''};
i.e. an energy barrier has to be overcome between the bound and unbound states \citep{BKZ67}. 
It is clear that in such cases, the additional energy $\Delta$
required to unbind the star is the magnitude of the energy barrier, not the (unphysical) $-W_0<0$.

The secular stability of a star against small adiabatic perturbations 
is not defined by the sign of the total energy $W$, but rather by the variational conditions:
an equilibrium state is an extremum of the total energy $(\delta W)_{\rm ad} =0$; and 
secularly stable configurations are then found at local minima: $(\delta^2 W)_{\rm ad} > 0$.
\citep{Chiu68}. 

If the first adiabatic exponent $\Gamma_1$ is approximately a constant throughout the star, 
the secular stability criterion reduces to the condition \citep{Chiu68}

\begin{equation}{
\Gamma_1 = \left ( \frac {\partial \ln P}{\partial \ln \rho } \right )_{\rm ad} > 4/3 
}\end{equation}
Here $P$ is the pressure and $\rho$ is the density.

For a one-zone model of a stellar envelope, a linearized version of the above condition,
the Baker's model \citep{Baker66} gives a similar criterion for the volume-averaged 
adiabatic exponent $\Gamma_1 > 4/3$ in the envelope 

On the other hand, the virial theorem, applied to the \emph{entire} star, 
gives us the condition for the total energy $W<0$ 
in the case of constant third adiabatic exponent $\Gamma_3$, 
as \citep[e.g.,][]{HKT04}:
\begin{equation}{
	\Gamma_3:=\left ( \frac{\partial \ln T}{\partial \ln \rho }\right )_{\rm ad} +1 > 4/3 \ .
}\end{equation}
Here $T$ is  is the temperature.
We note that this simple form depends crucially on Newton's third law applied to every pair of particles inside the star,
and thus needs to be substantially modified if we only consider the envelope.

We reiterate the importance of distinguishing these conditions: 
instability does not have to occur in a state with positive total energy, 
and $\Gamma_1$ does not have to coincide with $\Gamma_3$;
they are related as
\begin{equation}{
    	\Gamma_3=1 + \Gamma_1 \left ( \frac{\partial \ln T}{\partial \ln P }\right )_{\rm ad}
}\end{equation}
In particular, the ionization zones, where $
	\nabla_{\rm ad} = \left ( \frac{\partial \ln T}{\partial \ln P }\right )_{\rm ad} < 0.4
$
and can become as low as 0.1, have local value of $\Gamma_3<4/3$ while $\Gamma_1>4/3$.

It is unfortunately non-trivial to find the volume-averaged $\Gamma_1$ in the giant envelope
after depositing some heat in it.
However, it is clear that the criterion of the total energy in the envelope to be $W=0$
is not applicable for the envelope stability, nor is it a sufficient condition 
for the envelope to be dispersed to infinity. We will consider instead
the stability of a stellar envelope towards creating outflows.

\subsection{The condition for the envelope to outflow}

Since we are interested in the energy {\it requirements}, it is natural to consider 
a lower bound on $\Delta$: all orbital energy is converted into heat, and the velocity 
of ejecta at infinity is zero.
How precise these bounds are is not of great concern here and is a separate problem on 
its own, although we expect them to be reasonably 
close: the time scale for viscous friction between differentially rotating regions can exceed 
the spiral-in time, but most angular momentum is lost in the outer envelope while most of the energy
is released at closer orbits, so the overall effect of angular momentum conservation can at 
least be constructed to be small. 

As any other thermodynamical system undergoing a steady process, the material in a star obeys 
the first law of thermodynamics, which states that the change of the internal energy comes not 
only from the heat transferred into system, but also from the work done by the system:
\begin{equation} 
 \delta E = \delta Q - P \delta \left ( \frac{1}{\rho} \right ) \ .
\end{equation}
Applying this equation for each mass shell in the envelope, 
(see, e.g., \citealt{CW66, KS72}),
gives the energy conservation equation in Lagrangian coordinates as 

\begin{equation}
\frac{\partial}{\partial t} \left(\frac{u^2}{2} + \Psi +\epsilon\right )  +
  {P}\frac {\partial (1 / \rho) }{\partial t}  = 0 \ .
\end{equation}
Here  $u$ is the velocity. 

The initial condition is that at the start the mass shells in the envelope are not moving, 
but every shell in the envelope received some heat, in this case deposited from the companion's orbital contraction.
The heat has some arbitrary distribution ${\delta q(m)}>0$ in the envelope (${\delta q(m)}$ is per  mass unit), such that 

\begin{equation}
\int_{\rm core}^{\rm surface} \delta q(m) dm = Q.
\end{equation}

A given lagrangian shell, once it has been heated and started expansion, 
will have reached the point of no return in its expansion when its $\Sigma>0$ 
\citep[e.g.,][]{BKZ67,SK72} 

\begin{equation}
(\delta q(m) + \Psi + \epsilon + \frac{P}{\rho})_{\rm start} 
	= (\frac{1}{2} u^2 + \Psi + \epsilon  + \frac{P}{\rho})_{\rm exp}  
	= \Sigma = const
\label{bern}
\end{equation}
In this form, the equation can be also recognized  as version of the Bernoulli equation.

The quantity $\Sigma$, for stability analysis purposes, is more important than the total energy: 
bipolitropic stellar model that has $\Sigma < 0$ in its envelope but positive total energy 
is metastable as a whole, even though the run-away to infinity is energetically allowable; 
but a star with $\Sigma>0$ in its envelope will be always quasi-steadily outflowing \citep{BKZ67,BK02}. 
It also has been noted that it is a general feature of low-mass giants during double shell 
burning to establish $\Sigma>0$
in larger and large parts of their envelopes as they evolve (with He shell providing large energy 
inflow to the envelope with each He-shell flash), 
and could possibly being  responsible for envelope outflows and/or ejections with planetary nebula 
formation \citep{SK72}. 

In line with results of past stability analyses of when a star starts an outflow \citep[][and the references therein]{BK02}, we 
suggest that once a \emph{part} of the common envelope has obtained positive $\Sigma$, it will start outflowing, 
notwithstanding the sign of the envelope's total energy.
A stronger criterion would be to require the \emph{entire} envelope to have positive $\Sigma$.
Under assumption of the minimum energy requirement (velocity of gas at infinity is zero), 
we can write the energy conservation for a whole envelope as

\begin{equation}
Q + \int_{\rm core}^{\rm surface}\left (\Psi(m) +  \epsilon(m) + \frac{P(m)}{\rho(m)} \right ) dm  = 0 \ .
\end{equation}

The timescale for start such quasi-steady surface outflows is one on which the envelope redistributes the dumped heat,
i.e. the thermal timescale of the envelope - recall our earlier note about most energy being deposited in lower orbits. 
This time is about few hundred years \citep[e.g, it is about 1000 years for a $20 M_\odot$, see Fig.1 in ][]{Ivanova11}.
This makes the appearance of mass outflows  {\it natural} during the self-regulating spiral-in stage; 
such a stage could last for up to a thousand years \citep{podsi01};
but they likely will not take place if the common envelope event occurs on the dynamical time, as is the case 
e.g., in the case of a physical collision, or if the swallowed companion is too small to expand the giant envelope 
and establish a self-regulated slow spiral-in phase.

Note, that 

\begin{itemize} 
\item due to the presence of a non-negative term $P/\rho$, this condition occurs before 
the envelope's total energy become positive; 
\item the master equation (\ref{allam}) of the `$\lambda$-formalism` is a version of eq.(\ref{bern})
where all the velocities, as well as the work $P/\rho$ are simply neglected.
\end{itemize}

\noindent The term $P/\rho$ is of the order of magnitude of $\epsilon$;
for example, for an ideal gas, with no radiation pressure and ionization taken into account,
$P/\rho = {2}/{3}\, \epsilon$. The quantity $h=\epsilon + {P}/{\rho}$ is generally known as {\it enthalpy}.

In line with the classical `$\lambda$-formalism' and for ease of comparison, we introduce:

\begin{equation}
E_{h, \rm bind} = - \int_{\rm core}^{\rm surface} \left ( \Psi(m)+ h(m) \right )dm  = \frac {G m_1 m_{1,\rm e}} {\lambda_{\rm h} R_1}
\end{equation}

\begin{figure}[t]
  \includegraphics[height=.35\textheight]{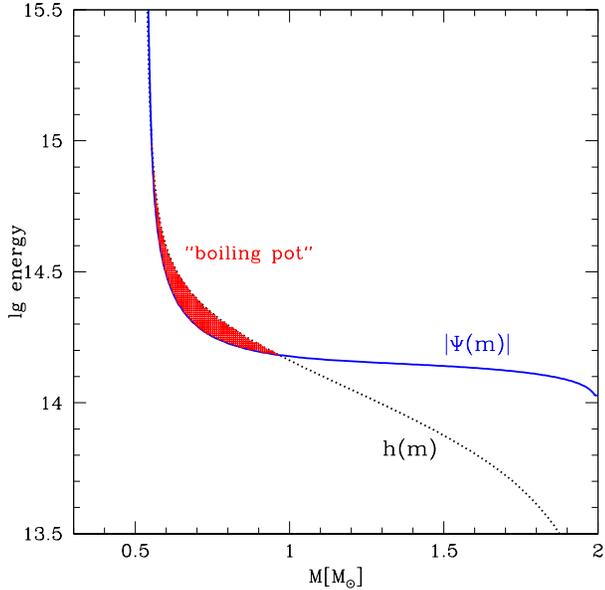}
  \caption{$\Psi$ and $h$ inside $2\ M_\odot$ giant. The ``boiling pot'' zone has excess energy. Energy is the specific energy in erg per g.
\label{bpz}
}
\end{figure}

Following \cite{SK72} in considering the detailed stellar models of giants with respect to their $\Sigma$, we also 
note that normal single giants have a "boiling pot'' zone (BPZ), where $\Sigma(m)=\Psi(m)+h(m) > 0$ 
due to high value of $P/\rho$  (Fig.~\ref{bpz}).
Once sufficient amount of the ``lid'' -- the star matter above this zone is removed, 
all material with positive $\Sigma$  could freely stream away without the need to convert any additional 
mechanical energy. 
The outer boundary of the BPZ almost does not
change during the giant stage, and the inner boundary does
not change once the convective envelope is established (e.g.,
in case of low-mass red giants, RGs). The bottom of the BPZ
roughly (but not always exactly) coincides with the bottom of
the outer convective zone. Illustratively, the mass contained
in this zone is: $\sim 0.5 M_\odot$ for $2 M_\odot$ RG, up to $7.7 M_\odot$ for a $30 M_\sun$ 
giant evolved without wind mass loss, and about 3 times less for a $30 M_\sun$ giant evolved with mass loss.
We anticipate that the presence of the boiling pot zone could be a key to understand why RGs
are expanding when losing their mass.

\section{Comparison of $E_{\rm \lambda, bind}$ and $E_{\rm h, bind}$ in stellar models}

Just like $\lambda$, $\lambda_{\rm h}$ can be found from detailed stellar calculations.
For the purpose of this comparison, 
we adopt the definition of the core mass for both cases as in \cite{Ivanova11}:
the post-CE core is the point with maximum  compression $P/\rho$ inside the Hydrogen burning shell, $m_{1, \rm cp}$.
Such a core will not re-expand after the envelope loss, and it was shown that in most of the cases 
it is energetically beneficial to remove the envelope down to exactly $m_{1, \rm cp}$.

In Table~1  we show the results for giants of different masses.
We used the set of stellar models described in \cite{Ivanova11}.
We find that in most giants the ratios of $\lambda_{\rm h}/\lambda$ are from $\sim2$ to $\sim5$, where
the largest ratios, for a giant of the same mass, are for an earlier giant.
The final separations allowed by the enthalpy-consideration are then larger by the same few times.

The minimal companion mass -- in the sense that no possible stripped core
value exists such that a binary would not merge -- changes by several times as well (Table~1).
Specifically, we want to point out the difference between the enthalpy consideration and standard prescription
for massive giants: while $\alpha_{\rm CE}\lambda$-formalism would predict the minimum surviving companion mass
of about several $M_\odot$, in line with the problem raised by \cite{Justham06}, the 
enthalpy-formalism allows for an low-mass companion to survive and form in a future a black hole LMXB.

\begin{table}
\caption{CE outcomes comparison. }
\begin{tabular}{l r  r r  r r    r r }
\hline
$m_{1} (m_{\rm zams})$ &$R_{1}$ & $m_{1,X}$  & $m_{1,\rm cp}$  & $\lambda$ & $\lambda_{\rm h}$  & $m_{2,\lambda}$ & $m_{2,h}$   \\
\hline
25.59(30) &  900  & 9.381  &  11.44 &  0.026  & 0.085    &  6.33 &  1.53  \\
25.53(30) &  1500  & 10.223  & 11.39      &  0.026  & 0.064     &    2.59 &  0.92  \\
18.5(20) &  600  & 5.59  & 6.48      &  0.065     &    0.299 &  2.84 &  0.46 \\
18.5(20) &  750  & 5.70  & 6.48      &  0.133     &    0.309 &  0.82 &  0.32 \\
16.8(20) &   850  & 6.75  & 6.92       &  0.067  & 0.142     &    0.72 &  0.31 \\
 9.75(10) &  200  & 1.69  & 1.95      &  0.148  & 0.274     &    1.87 &  0.86 \\
 9.75(10) &  300  & 1.73  & 2.04      &  0.136  & 0.244     &    1.28 &  0.62  \\
 9.74(10) &  360  & 1.95  & 2.10      &  0.143  & 0.253     &    0.74 &  0.37  \\
 5.09(10) &  380  & 2.87  & 2.94      &  0.061  & 0.109     &    0.16 &  0.09  \\
 4.99(5) &  40  & 0.575  & 0.725      &  0.402  & 0.815     &    1.9 &  0.75  \\
4.99(5) &  80  & 0.702  & 0.784      &  0.425  & 0.822     &    0.56 &  0.25  \\ 
2 &    10  & 0.253  & 0.271      &  1.167  & 2.804     &    0.39 &  0.13   \\
2  &  40  & 0.526  & 0.529      &  0.730  & 1.652     &    0.04 &  0.02    \\
1 &  10  & 0.253  & 0.254      &  0.941  & 2.29     &    0.04 &  0.02   \\

\hline
\end{tabular}
\label{table}

\footnotesize{A post-CE mass is adopted to be the divergence point $m_{1,\rm cp}$, for a comparison is shown the mass of
the hydrogen-exhausted core $m_{1,\rm X}$ (where $X\le 10^{-10}$).
$\lambda$ and  $\lambda_{h}$ connect the energies required to eject the envelope with their parametrizations in two formulations (eqs.~2 and 12).
$m_{2,\lambda}$ and  $m_{2,h}$ are the {\it minimum} companions' masses
that could survive a CE event, where $m_{2,\lambda}$ is using standard prescription and
$m_{2,h}$ is with enthalpy-consideration.
$m_1$ are the current donor masses and $m_{\rm zams}$ are donor  masses at the zero-age main sequence (if different from $m_1$).
$R_1$ are the current donor radii.
All masses are in $M_\odot$, $R_{1}$ is in $R_\odot$.
}
\end{table}

\section{Conclusions}

In this {\it Letter}  we  considered the termination of a common envelope event at the moment it establishes
a quasi-stationary mass outflow that would reach a point of no return.
Such outflow develops during the slow self-regulating spiral-in phase of the common envelope event.
We showed that if the common envelope is escaping the binary in this way,
neglecting the ${P}/{\rho}$ term in the standard $\alpha_{\rm CE}\lambda$ energy 
conservation prescription is too gross.
If the enthalpy rather than internal energy 
is calculated in the energy balance, it makes a crucial difference 
in giants.

While we anticipate that the estimate of a proper energy requirement is paramount
for the formation rates of all kinds of post-CE binaries,
we especially emphasize the importance of this effect in the case of CE with a massive giant.
When the standard energy formalism predicts that only a several solar mass donor could survive CE,
and so the formation of a low-mass X-ray binary with a black hole accretor is forbidden unless
$\alpha_{\rm CE}$ exceeds 1, 
the enthalpy-consideration naturally allows for a low-mass companion survival.

\section{Acknowledgment}
Natalia Ivanova acknowledges support from NSERC and Canada Research Chairs Program.
We are thankful to Jonathan Braithwaite for spotting a misprint in equation (8) of the version 1 of this paper.

\end{document}